\documentclass[sigplan,10pt]{acmart}%,anonymous,review
\acmConference[]{August}{14}{2020}
\acmYear{Preprint}
\acmISBN{} % \acmISBN{978-x-xxxx-xxxx-x/YY/MM}
\acmDOI{} % \acmDOI{10.1145/nnnnnnn.nnnnnnn}
\startPage{1}

\setcopyright{none}
\usepackage{booktabs}   %% For formal tables:
\usepackage{subcaption} %% For complex figures with subfigures/subcaptions
\usepackage{tabularx}
\usepackage{listings}
\usepackage[vlined,linesnumbered,ruled]{algorithm2e}
\usepackage{footnote}
\usepackage[subtle]{savetrees}

\newcommand{\TT}[1]{\texttt{#1}}
\newcommand{\BF}[1]{\textbf{#1}}

\newcommand{\REM}[1]{}

\definecolor{dmlgreen}    {RGB}{51,  160,  44}
\definecolor{dmlblue}     {RGB}{31,  120, 180}
\definecolor{dmlred}      {RGB}{202,   0,  32}
\usepackage{enumitem}

\begin{document}

%% Title information
\title[Systolic GPU]{Systolic Computing on GPUs for Productive Performance}

%% Author information
%% Contents and number of authors suppressed with 'anonymous'.
%% Each author should be introduced by \author, followed by
%% \authornote (optional), \orcid (optional), \affiliation, and
%% \email.
%% An author may have multiple affiliations and/or emails; repeat the
%% appropriate command.
%% Many elements are not rendered, but should be provided for metadata
%% extraction tools.

%% Author with single affiliation.
\author{Hongbo Rong}
\affiliation{
  \institution{Intel}
}
\email{hongbo.rong@intel.com}

%% Author with two affiliations and emails.
\author{Xiaochen Hao}
\affiliation{
  \institution{Intel, Peking University}
}
\email{xiaochen.hao@intel.com}         %% \email is recommended

\author{Yun Liang}
\affiliation{
  \institution{Peking University}
}
\email{ericlyun@pku.edu.cn}

\author{Lidong Xu}
\affiliation{
  \institution{Intel}
}
\email{lidong.xu@intel.com}

\author{Hong H Jiang}
\affiliation{
  \institution{Intel}
}
\email{hong.h.jiang@intel.com}

\author{Pradeep Dubey}
\affiliation{
  \institution{Intel}
}
\email{pradeep.dubey@intel.com}

\begin{abstract}

We propose a language and compiler to productively build high-performance {\it software systolic arrays} that run on GPUs. Based on a rigorous mathematical foundation (uniform recurrence equations and space-time transform), our language has a high abstraction level and covers a wide range of applications. A programmer {\it specifies} a projection of a dataflow compute onto a linear systolic array, while leaving the detailed implementation of the projection to a compiler; the compiler implements the specified projection and maps the linear systolic array to the SIMD execution units and vector registers of GPUs. In this way, both productivity and performance are achieved in the same time. This approach neatly combines loop transformations, data shuffling, and vector register allocation into a single framework. Meanwhile, many other optimizations can be applied as well; the compiler composes the optimizations together to generate efficient code.

We implemented the approach on Intel GPUs. This is the first system that allows productive construction of systolic arrays on GPUs. We allow multiple projections, arbitrary projection directions and linear schedules, which can express most, if not all, systolic arrays in practice. Experiments with 1- and 2-D convolution on an Intel GEN9.5 GPU have demonstrated the generality of the approach, and its productivity in expressing various systolic designs for finding the best candidate. Although our systolic arrays are purely software running on generic SIMD hardware, compared with the GPU's specialized, hardware samplers that perform the same convolutions,  some of our best designs are up to 59\% faster. Overall, this approach holds promise for productive high-performance computing on GPUs.
\end{abstract}

%% 2012 ACM Computing Classification System (CSS) concepts
%% Generate at 'http://dl.acm.org/ccs/ccs.cfm'.


  %% \keywords are mandatory in final camera-ready submission

\maketitle

\section{Introduction}
\label{sec:intro}
All modern GPUs achieve performance via hardware multi-threading and SIMD (single-instruction multiple-data), and an efficient memory hierarchy~\cite{closerLookAtGPU}.  The mainstream programming languages such as CUDA and OpenCL essentially expose an SIMT (single-instruction multiple-threads) programming interface, and rely on an underlying compiler to transparently map a wrap of threads to SIMD execution units. If data need to be exchanged among threads in the same wrap, programmers have to write explicit shuffle instructions~\cite{cudaProgrammingGuide}.

This paper proposes a new programming style that programs GPUs as building {\it software systolic arrays}. Systolic arrays have been extensively studied since 1978~\cite{KungLeiserson78}, and shown an abundance of practice-oriented applications, mainly in fields dominated by iterative procedures~\cite{algo-informatics-online}, e.g. image and signal processing, matrix arithmetic, non-numeric applications, relational database~\cite{whySystolic,DongarraQR2013, cnn_systolic, baseline-gemm, transitive_closure, pairhmm-intel, HugheyPhdThesis91}, and so on.

A {\it systolic array} is composed of many hardware {\it PEs} (Processing Elements) that have the same instructions and works rhythmically: every time step, the PEs typically read inputs from some neighbors, process the inputs, and forward the inputs or results to other neighbors. Therefore,  systolic arrays can be viewed as ``the combination of SIMD and the pipeline architectures characteristics"~\cite{IntroSystolic}, a fact that has been observed for a long time.

Based on this fact, we can build a systolic array on a GPU by mapping the PEs to the SIMD lanes of the GPU, and realizing the data forwarding using shuffle instructions among vector registers. This idea has been proposed as ``software systolic array" by Chen et al., and demonstrated on stencils and convolution manually with competitive performance~\cite{Chen19SoftSystolic}. Similar ideas can be found elsewhere. For examples, Pondemente, Luna and Alba used GPUs to build systolic arrays for a generic search algorithm~\cite{systolicOptGPU}, Wang et al. implemented sequence alignment algorithms on GPUs in a systolic style without mentioning it~\cite{WangjieSequenceAlignment}. All these works were done on Nvidia GPUs in the CUDA language.

These state-of-art works, however, focus on specific workloads, and the systolic arrays were built in high programming skills. None of the works has pointed out a general, systematic solution how to build on GPUs arbitrary systolic arrays for the numerous workloads that are known to benefit from systolic algorithms. And they have not provided a productive tool for quickly building  systolic arrays on GPUs, either.

In this paper, we present a language and compiler to productively build high-performance systolic arrays that run on GPUs. A programmer {\it specifies} a dataflow compute in {\it uniform recurrence equations} (UREs), and a projection of the compute onto a linear systolic array in a {\it space-time transform}. We allow multiple projections, arbitrary projection directions and linear schedules. Our approach has the following characteristics:

\begin{enumerate}
    \item Generality.
    
UREs and space-time transform are the theoretical foundation of most systolic arrays we can see in the real world. They are rigorously formulated in mathematics and have been extensively studied. By enabling them, our language and compiler are able to cover the wide range of applications to which  numerous systolic arrays apply.
    
    \item Productivity.
    
Our language has a higher abstraction level than the popular, SIMT-based programming languages like CUDA and OpenCL. Both UREs and space-time transforms are succinct math: UREs are a functional notation of a compute, and space-time transforms are expressed by matrices.    

Our language separates concerns of programmers. For a compute, programmers write its functional notation (UREs) and its optimizations (space-time transforms, and other loop and data optimizations) separately. 

Our language is a specification language: programmers only specify optimizations, but leave their detailed implementation to a compiler; the compiler implements the optimizations, particularly, maps linear systolic arrays determined by the specified space-time transforms to the SIMD execution units and vector registers of GPUs. In this way, both productivity and performance are achieved in the same time. 

    \item Performance.

A space-time transform neatly combines loop transformations, data shuffling, and vector register allocation together: it can transform a loop nest in a single shot with the combined effect of several loop transformations (like loop reordering, skewing and vectorization), allocation of minimum number of vector registers, and shuffling of the data in the registers.  

Meanwhile, many other optimizations can be applied as well. These optimizations include tiling, multi-threading, building a memory hierarchy with shared memory and registers, etc. The compiler composes the optimizations together to generate efficient code.

\end{enumerate}

We prototyped a system for our approach on Intel GPUs~\cite{Gen9Intro}. We leverage Susy~\cite{Susy}, a system that enabled UREs and limited space-time transforms for FPGAs. Susy allows only a single projection, and the projection direction must follow a loop dimension. In our system, we break the limitations to allow multiple projections and arbitrary projection directions, and generate code for GPUs, which is essentially different from FPGAs, and necessitates {\it substantial innovation}. From the perspective of programming style, Susy for FPGAs is single-thread SIMD, while our language for GPUs is a mix of SIMT and SIMD. From the perspective of hardware architectures, FPGAs have on-chip memory but no hardware caches, have massive programmable logic resources and registers, while GPUs have shared memory, hardware caches, fixed execution units, and limited thread-private registers, and thus the compiler has to optimize very differently.  

\begin{figure}[!th]
    \centering
    \includegraphics[width=\linewidth]{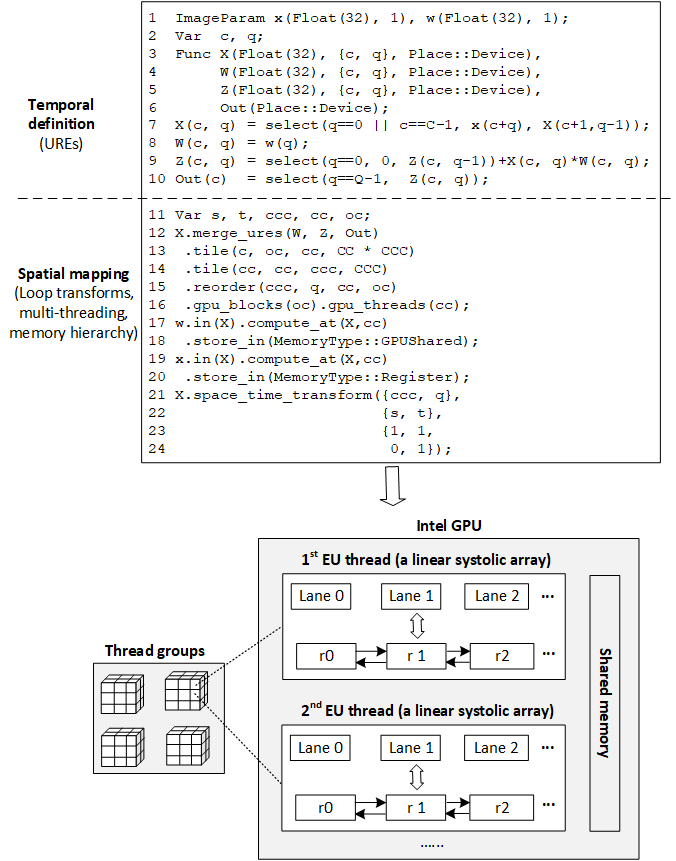}
    \caption{The overall flow.  }
    \label{fig:T2S-GPU-overall-arch}
\end{figure}

Fig.~\ref{fig:T2S-GPU-overall-arch} provides a high-level overview of our system. Programmers write a specification, which contains two parts: (1) a {\it temporal definition}, which is a functional definition of the compute to perform, in the form of UREs, and (2) a {\it spatial mapping}, which is optimizations to map the compute efficiently to a GPU. These optimizations transform loops (e.g. by loop tiling, reordering, space-time transform, etc.), enable multi-threading, and build a memory hierarchy. 
In more detail, the iteration space of a dataflow compute is cut into groups of software threads, and scheduled group by group to a GPU at runtime; every software thread is mapped to a hardware thread called {\it Execution Unit} (EU) thread. For brevity, we will use the term {\it thread} to refer to a software thread, and {\it EU thread} for its corresponding hardware thread.

Unlike a traditional thread, a thread here contains a linear systolic array, as specified by a space-time transform as part of the spatial mapping. The linear systolic array is realized by the SIMD function units and vector registers in a EU thread. More specifically, every lane of a SIMD function unit is a PE, and the PEs exchange data by shuffling the vector registers that contain the data. One may view a thread here as a wrap of threads in CUDA.

This is the first system that allows productive construction of  systolic arrays on GPUs, to the best of our knowledge. We remark that although the current system targets only Intel GPUs, it is general and should be applicable to other GPUs as well.

We performed initial experiments with 1- and 2-dimensional convolution on an Intel Gen9.5 GPU, and compared the performance with the hardware samplers of the GPU performing the same convolutions. The experiments show that the approach is very general and flexible; various systolic designs of the convolutions can be expressed succinctly in 10-20 minutes, and thus programmers can quickly search for the best designs; the best designs are 14\%-59\% faster than the specialized hardware samplers, which is very encouraging since our systolic arrays are purely software on generic SIMD hardware. Overall, these initial results show that our approach holds promise for productive high-performance computing on GPUs.

\section{Background}
\label{sec:background}

In this section, we briefly introduce necessary background knowledge. Throughout the paper, {\it any vector is assumed to be a column vector, except pointed out otherwise}.

\subsection{Uniform Recurrence Equation}
\label{sec:ures}

We follow Quinton's definition of URE~\cite{Quinton:1984:URE} with slight generalization~\cite{Xue:1992:FSC}. A system of UREs are equations,  each of which has the following form: 
\begin{equation}
    V_i(\vec{z}) = f(V_1(\vec{z}-\vec{e_1}), V_2(\vec{z}-\vec{e_2}), ..., V_p(\vec{z}-\vec{e_p)}) \label{equ:ure}
\end{equation}
where $V_1, V_2, ..., V_p$ are {\it variables}, $f$ is an arbitrary function, $\vec{z}$ is a {\it (computation) point} (i.e. an iteration) in an $n$-dimensional space, and point $\vec{z}$ reads variables  $V_1, V_2, ..., V_p$ from the previous points at constant distances of $\vec{e_1}, \vec{e_2}, ..., \vec{e_p}$, respectively ({\it We assume that there is only one dependence associated with one variable}). In other words, the UREs represent an $n$-deep loop nest with a uniform dependence structure.

UREs are a functional way to express a computation so that any memory location $V_i(\vec{z})$ is written only once. This is called  dynamic single-assignment (DSA)~\cite{Vanbroekhoven:2007:PDS:1278349.1278353}. The read-write relationship between computation points is explicit, which makes it easy for a compiler to analyze and parallelize the computation. 

\subsection{Space-time Transform}
\label{sec:space-time-transform}
\begin{figure}[!tb]
    \centering
    \includegraphics[width=\linewidth]{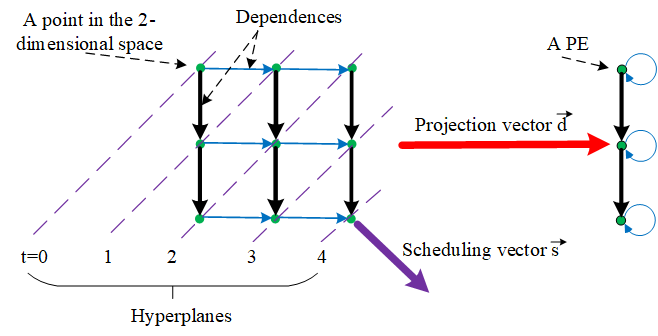}
    \caption{Illustrating the concept of space-time transform. Here 9 points in a 2-dimensional space are projected to run on 3 PEs in a 1-dimensional space. The dependences are projected accordingly.}
    \label{fig:stt-concept}
\end{figure}

A {\it space-time transform} projects the computation points in an $n$-dimensional space to an  $n-1$-dimensional space, and schedules the points to execute in temporal order. Following the notations of Chen and Kung~\cite{multiProjectionKung}, let $\vec{d}$ be a {\it projection vector} that projects the points to the $n-1$-dimensional space. In the $n-1$-dimensional space, the coordinates of the points are $n-1$-dimensional vectors, and can be viewed as linear combinations of the column vectors of a $(n - 1) *n$  {\it allocation matrix} (also called {\it projection matrix}) $\mathbf{P}$.  The projection vector is a  normal vector of the $n-1$-dimensional space and thus there must be $\mathbf{P}\vec{d}=0$. Also let $\vec{s}$ be a {\it scheduling vector} representing the time scheduling.
Then  a point $\vec{z}$ in the original space is mapped to  
$\bigl( \begin{smallmatrix} 
  \mathbf{P}\\
  \vec{s}^T
\end{smallmatrix} \bigr)\vec{z}=\bigl( \begin{smallmatrix} 
  \mathbf{P} \vec{z}\\
  \vec{s}^T \vec{z}
\end{smallmatrix} \bigr)$. 
That is, point $\vec{z}$ in the original space is allocated to point $\mathbf{P}\vec{z}$ in the new space, and is scheduled to execute at time step $\vec{s}^T \vec{z}$. The points in the new space is what we referred to as PEs before.
See Fig.~\ref{fig:stt-concept} for an illustration.

A transform is valid if and only if the following conditions are satisfied~\cite{multiProjectionKung}: 
\begin{itemize}
    \item {\it Data availability:} $\vec{s}^T \vec{e} \ge 0$ for any dependence $\vec{e}$. $\vec{s}^T \vec{e} \neq 0$ for non-broadcast data.
    \item {\it Processor availability:} $\vec{s}^T \vec{d} > 0$.
\end{itemize}
The first condition ensures that for any dependence (except read-after-read, or so-called broadcast), its source point has to be executed before its sink point. The second condition ensures that two computation points, if mapped to the same PE, do not execute at the same time.

\begin{figure*}[!tb]
    \centering
    \includegraphics[width=\linewidth]{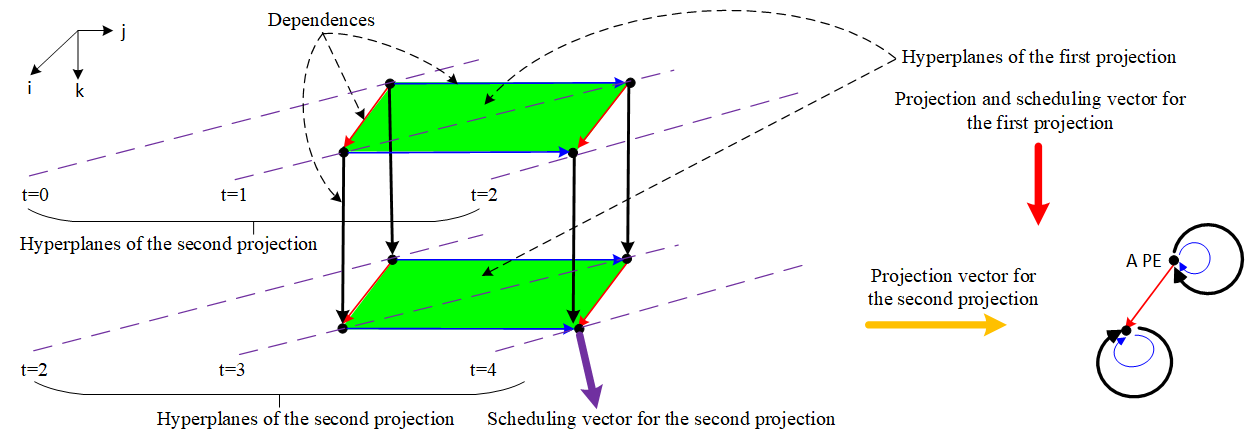}
    \caption{Illustration of multiple projection. In this example, there are 2 projections that project a 3-dimensional space $(i, j, k)$ into a 1-dimensional space. 
    When expressing the two projections in a single space-time transform, one may find that the time schedule equals $i + j + 2*k$; in other words, the scheduling vector $\vec{s}$ equals $\bigl(\begin{smallmatrix} 
  1 & 1 & 2
\end{smallmatrix} \bigr)^T$. One may also find that any compute point $(i, j, k)$ are mapped to a PE indexed by $i$, i.e. the allocation matrix equals $\bigl( \begin{smallmatrix} 
  1 & 0 & 0
\end{smallmatrix} \bigr)$. }
    \label{fig:mutli-projection-concept}
\end{figure*}

{\it Multiple projections} can be applied to an $n$-dimensional space. They can be collectively expressed by a single space-time transform. Fig.~\ref{fig:mutli-projection-concept} illustrates the concept with a 3-dimensional space projected to a 1-dimensional space after two projections. This example, however, can be generalized to arbitrary multiple projections. We leave the detailed theory to a literature~\cite{multiProjectionKung}.

\section{Language and Programming Model}
\label{sec:language}

As we said, our language is a specification language for GPUs. Typically, a specification is written in our language following a specific pattern: a set of UREs expressing a temporal definition, followed by a spatial mapping. In the spatial mapping, all UREs are merged together as the body of the same loop nest; for this loop nest, loop tiling and reordering are usually performed first, so as to create a final shape of the loop nest; then some outermost loops are designated as the loops for scanning thread groups, and for scanning the threads in a group, respectively; after that, the input data are loaded either to shared memory (and then into registers automatically by the compiler), or to registers directly; finally, other loop optimizations are applied for performance; particularly, every thread is built into a systolic array by a space-time transform of the UREs. In our programming model, registers are private to threads.

Table~\ref{table:language-features} describes a minimal set of the language features. The \TT{space\_time\_transform} primitive is the main feature this paper proposes. As said before, it is much more general than the same named feature in Susy~\cite{Susy}, and internally targets GPUs instead of FPGAs. In a space-time transform, we allow only linear schedules, which is the most popularly used in practice. Thus this is not a real restriction in the real world. 

\begin{table*}[h]
  \centering
  \setlength{\tabcolsep}{3pt}
  \captionsetup{belowskip=8pt,aboveskip=4pt}
  \caption{Primitives of our language.}
  \footnotesize%\small
  \begin{tabularx}{\linewidth}{l X}
    \toprule
    \BF{Primitive} & \BF{Description} \\
    \midrule
    \TT{Func F([return type, arguments, place])} & Declare a function. The \TT{return type} and \TT{arguments} must be declared for a recurrence function. Place is where the function will be running, either CPU (i.e. \TT{Place::Host}) or GPU (i.e. \TT{Place::Device}). The default place is CPU.\\
    \midrule
    \TT{F(arguments)=select(condition, value1 [, value2])} & Define a function. The function returns \TT{value1} if \TT{condition} is true, or \TT{value2} otherwise. If \TT{value2} is not specified, the function is undefined when \TT{condition} is false. \\
    \midrule
    \TT{F.merge\_ures(G, H, ...) } & Put the set of UREs \TT{F, G, H, ...} as the body of the loop nest of function \TT{F}. \\
    \midrule
    \TT{F.tile(var, var$_\TT{o}$, var$_\TT{i}$, factor)} & Tile the loop variable \TT{var} of \TT{F} into two new variables \TT{var$_\TT{o}$} and \TT{var$_\TT{i}$} with a factor of \TT{factor}. \\
    \midrule
    \TT{F.reorder(var$_1$, var$_2$, ..., var$_\TT{n}$)} & Reorder the loop nest of function \TT{F} according to the specified order (starting from the innermost level). \\
        \midrule
    \TT{F.gpu\_blocks(x, [y, [,z]])}  & The groups of threads are scanned by loop  \TT{x} (and \TT{y, z} if specified). \\
    \midrule
    \TT{F.gpu\_threads(x, [y, [,z]])} & The threads in a group are scanned by loop  \TT{x} (and \TT{y, z} if specified).\\
    \midrule
    \TT{F.in(G)} & The call to \TT{F} in function \TT{G}. \\
    \midrule
    \TT{F.compute\_at(G, x)} & Compute function \TT{F} at loop level \TT{x} of function \TT{G}, which calls function \TT{F}. \\
    \midrule
    \TT{F.store\_in(memory type)} & Store the values of function \TT{F} in a given type of memory, e.g. shared memory or registers.\\
    \midrule
        \TT{F.space\_time\_transform(source loops, } & Map the  \TT{source loops} of function \TT{F} to the \TT{destination loops} with the \TT{transform matrix}.\\ 
    \TT{   destination loops, transform matrix} & The loops are specified to start from the innermost level. The \TT{transform matrix} is in the\\
    \TT{   [, reverse transform]               }  &  form of $\bigl( \begin{smallmatrix} 
  \mathbf{P}\\
  \vec{s}^T
\end{smallmatrix} \bigr)$, where $\mathbf{P}$ and $\vec{s}$ are the allocation matrix and scheduling vector, respectively
\\
    \TT{   [, SpaceTimeTransform::CheckTime])}  &  (See Section~\ref{sec:background}). \\
&    The \TT{reverse transform} describes how to map from the destination loop variables after the transform back to the source variables before the transform. The \TT{reverse transform} is optional when the \TT{transform matrix} is unimodular, in which case the compiler can automatically compute the reverse transform. \\
&    If \TT{SpaceTimeTransform::CheckTime} is specified, the compiler  generates an explicit check for each PE to execute only when its time steps come; otherwise, every PE always runs and the programmer needs ensure that does not affect correctness.\\
    \bottomrule
  \end{tabularx}
  \label{table:language-features}
  %\vspace{-2mm}
\end{table*}

\subsection{Illustrating the Language, Programming Model and IR Changes}
\label{sec:spec-for-1-d-convolve-SBM}

Fig.~\ref{fig:T2S-GPU-overall-arch} illustrates our language with a specific design for 1-D convolution. The mathematical definition of 1-D convolution is
\begin{equation}
      Z(c) = \sum\limits_{q}x(c+q) * w(q)
      \label{equ:1d-convolve}
\end{equation}
where $x$ is a sequence of input data, and $w$ is a short sequence of weights. At every position $c$ of the output sequence, the weight sequence multiplies with the corresponding input data, and generates one result $Z(c)$. 

There are many systolic designs for convolution accelerators. According to Jo, Kim and Park~\cite{JoKimPark2018}, these designs can be classified based on the following ordered combinations of inputs, weights and partial sums: 

An input or weight 
\begin{itemize}[topsep=0pt]
    \item B: is Broadcast to all PEs, 
    \item F: is Forwarded from one PE to another, or
    \item S: Stays in a PE,
\end{itemize}
and a partial sum is 
\begin{itemize}[topsep=0pt]
    \item A: Aggregated from all PEs, 
    \item M: Migrated from one PE to another, or 
    \item S: Sedimented in a PE.
\end{itemize}

\begin{figure}[!tbh]
\includegraphics[width=\linewidth]{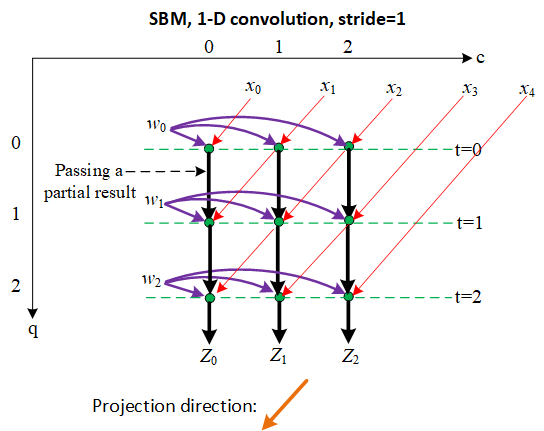} 
\caption {SBM systolic design for 1-D convolution. }
\label{fig:1-d-conv-SBM}
\end{figure}

Fig.~\ref{fig:1-d-conv-SBM} illustrates one of the designs, namely SBM~\cite{JoKimPark2018}, in which the inputs stay in each PE, the weights are broadcast to all the PEs, and the partial sums migrate from a PE to another.

The top half of Fig.~\ref{fig:T2S-GPU-overall-arch} contains a specification to express this design. The capitalized symbols in the example specification (and in the rest of the paper) like \TT{CC} etc. represent static constants. In the temporal definition of the specification,  Line 7-9 define the 1-D convolution according to Equation~\ref{equ:1d-convolve}, but in a recurrent form. These recurrent equations can be intuitively derived by following the dataflow in Fig.~\ref{fig:1-d-conv-SBM}.  Line 10 defines an output function \TT{Out} that takes a final value of \TT{Z} after the corresponding recurrent computation is done. Function \TT{Out} is not really a URE, but is treated so by the compiler to be simple. As we can see, the temporal definition of a compute is close to the original mathematical definition in Equation~\ref{equ:1d-convolve}, and exposes the dataflow. This makes it easy for programmers to define the functionality of the compute and for the compiler to analyze dependences, which is an advantage of the programming model.

In the spatial mapping, Line 12 makes the UREs the body of the loop nest of function \TT{X}. The compiler will build an {\it intermediate representation} (IR) for the loop nest like this:

\begin{lstlisting}[
    language=C,numbers=none,stepnumber=1,showstringspaces=false,tabsize=1,breaklines=true,escapechar=`]
    for (c = 0; c < C; c++)
     for (q = 0; q < Q; q++)
      X(c,q) = select(q==0||c==C-1,x(c+q), X(c+1,q-1))
      W(c,q) = w(q)
      Z(c,q) = select(q==0, 0, Z(c,q-1))+X(c,q)*W(c,q)
      Out(c) = select(q==Q-1,  Z(c,q))
\end{lstlisting}

Then Line 13-15 split the outer loop twice into 3 loops, and  reorder all the loops. Then Line 16 designates the outermost two loops for scanning thread groups and threads, respectively. Line 17-20 load the weights and inputs for each thread into shared memory and registers, respectively. 
Consequently, the loop nest becomes like this:

\begin{lstlisting}[
    language=C,numbers=none,stepnumber=1,showstringspaces=false,tabsize=1,breaklines=true,escapechar=`] 
    parallel for (oc=0; oc< C/CC/CCC; oc++) // Groups
     parallel for (cc = 0; cc < CC; cc++)   // Threads
      if (oc == 0 && cc == 0) 
        allocate, and load weights into, shared memory
      load weights from shared memory into registers
      load inputs to be used into registers
      for (q = 0; q < Q; q++)
       for (ccc = 0; ccc < CCC; ccc++)
        c = oc * CC * CCC + cc * CCC + ccc
        X(c,q)=select(q==0||c==C-1,x(c+q),X(c+1,q-1))
        W(c,q)=w(q)
        Z(c,q)=select(q==0,0, Z(c,q-1))+X(c,q)*W(c,q)
        Out(c)=select(q==Q-1, Z(c,q))
\end{lstlisting}

Line 21-24 express the systolic array determined by the projection shown in Fig.~\ref{fig:1-d-conv-SBM}. The source loops \TT{ccc} and \TT{q} are mapped to a space loop \TT{s} and a time loop \TT{t} by a transform matrix $\bigl(\begin{smallmatrix} 
  1 & 1\\
  0 & 1
\end{smallmatrix} \bigr)$, which means that \TT{s = ccc + q} and \TT{t = q}. This is a unimodular matrix, and the compiler can automatically find the reverse transform. The compiler  maps variable \TT{X}, \TT{W}, and \TT{Z} to vector registers based on the transform matrix, and determines the sizes of the vector registers as \TT{Q + CCC -1}. The compiler then replaces the references to the variables  with references to the vector registers. Consequently, the two source loops \TT{ccc} and \TT{q} become like this:

\begin{lstlisting}[
    language=C,numbers=none,stepnumber=1,showstringspaces=false,tabsize=1,breaklines=true,escapechar=`] 
      vector<float, Q + CCC - 1> X, W, Z; 
      for (t = 0; t < Q; t++)
        vectorize for (s = 0; s < Q + CCC - 1; s++)
          q, ccc = t, s - t      // Reverse transform
          c = oc * CC * CCC + cc * CCC + ccc
          X(s)=select(q==0||c==C-1,x(c+q),X(s))
          W(s)=w(q)
          Z(s)=select(q==0,0, Z(s-1))+X(s)*W(s)
          Out(c)=select(q==Q-1, Z(s))
\end{lstlisting}

In the next section, we will explain in more detail how the above evolution of IR is realized by the compiler.

\section{Compiler}
\label{sec:compiler}

\begin{figure}[!t]
    \centering
    \includegraphics[width=0.8\linewidth]{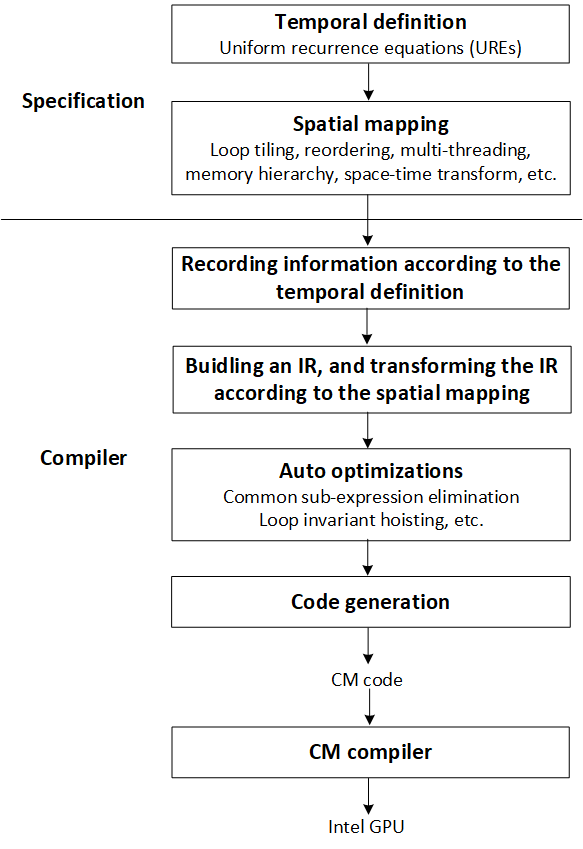}
    \caption{The compilation flow.}
    \label{fig:compile-flow}
\end{figure}

Fig.~\ref{fig:compile-flow} shows our compilation flow. As we said, a specification contains a temporal definition and a spatial mapping. The compiler first records information according to the temporal definition, then based on that information, building an IR and transforming the IR according to the spatial mapping. This is a {\it reactive compilation phase}, in which the compiler performs only the specified optimizations. After that, the compiler enters a {\it proactive compilation phase}, where the compiler transparently optimize the IR. For example, the compiler may perform common sub-expression elimination to reduce the strength of computation, hoist loop invariant out of a loop, etc. Finally, the compiler generates code for the target GPU from the IR. Currently, we generate GPU code in the CM language~\cite{CM}, which extends the standard C++ language with explicit SIMD support for GPUs to exploit data parallelism in applications. Particularly, our linear systolic arrays are realized in the CM language using matrices/vectors and SIMD operations that manipulate the matrices/vectors. Finally, we invoke the CM compiler to generate binaries for Intel GPUs. 

\subsection{Space-time Transform}
\label{sec:codegen-space-time-transform}
Here we describe the compiler implementation of space-time transform in more detail. As we said, no matter we project a loop nest once or multiple times, we can express the projection(s) in a single space-time transform. Let $\mathbf{P}$ be the allocation matrix, and $\vec{s}$ be the scheduling vector.
To be simple but without losing generality, let us say the original IR is

\begin{lstlisting}[
    language=C,numbers=left,stepnumber=1,showstringspaces=false,tabsize=1,breaklines=true,escapechar=`] 
        $\mathbf{Code\ 1:\ Original\ IR\ Example}$
    int $V$[extents of $\vec{z}$];
    for loops of $\vec{z}$
      $V[\vec{z}] = f(V[\vec{z} - \vec{e}]) + g(\vec{z})$
\end{lstlisting}
where $V$ is a recurrent variable with a dependence distance vector $\vec{e}$, and $f$ and $g$ are arbitrary functions. Then in general, the compiler transforms the IR as follows:

\begin{lstlisting}[
    language=C,numbers=left,stepnumber=1,showstringspaces=false,tabsize=1,breaklines=true,escapechar=`]
        $\mathbf{Code\ 2:\ Space-time\ Transformed\ IR\ Example}$
    int $V$[extent of $\mathbf{P}\vec{z}$][$\vec{s}^T\vec{e}$ + 1];    
    for t = min($\vec{s}^T\vec{z}$); t <= max($\vec{s}^T\vec{z}$); t++
     vectorize for each PE $\vec{x} \in \mathbf{P}\vec{z}$
       $V[\vec{x}][1, 2, ..., \vec{s}^T\vec{e}$] = $V[\vec{x}][0, 1, ..., \vec{s}^T\vec{e}$-1]
       $\vec{z} = h(\vec{x}, t)$
       if $\vec{z}$ is in the original iteration space
          $V[\vec{x}][0] = f(V[\vec{x} + \mathbf{P}\vec{e}][\vec{s}^T\vec{e}]) + g(\vec{z})$
\end{lstlisting}
Here \TT{h} is a function that reverses the space-time transform so that from the current PE and time step, we can find the corresponding computation point in the original iteration space before the transform. If the transform matrix is unimodular, the compiler can compute its inverse matrix and that is the reverse transform \TT{h}. Otherwise, the programmer must specify the reverse transform.

Note that in the above transformed IR, variable \TT{V} in Line 2 must be a matrix: (1) $\mathbf{P}\vec{z}$ is the index of the PE on which  the original point $\vec{z}$ is to run; the index is a scalar as the transform is for building a linear systolic array. Therefore, the first dimension of \TT{V}, the extent (i.e. the upper bound) of $\mathbf{P}\vec{z}$ is the number of PEs in the linear systolic array, and is a static constant. (2) $\vec{s}^T\vec{e}$ is constant given that both $\vec{s}$ and $\vec{e}$ are static constant vectors. Therefore,  the second dimension of \TT{V} is also a static constant. A special case is that when $\vec{s}^T\vec{e}$ equals 0, variable \TT{V} degenerates into a vector. 

Therefore, in generating code for the transformed loop nest, the compiler maps variable \TT{V} in Line 2 above to a matrix or vector. The dimensions of the matrix or vector are determined by the allocation matrix $\mathbf{P}$ and scheduling vector $\vec{s}$, in other words, {\it space-time transform also performs register allocation for recurrent variables}. In Line 5, every PE shuffles its own data. In Line 8, every PE gets an input from a neighbor PE at a constant space distance ($\mathbf{P}\vec{e}$) and a constant time distance ($\vec{s}^T\vec{e}$), performs some compute locally, and saves the results locally. Thus we can see that the entire body of the loop nest can be vectorized. That is why the loop in Line 4 is annotated as vectorized.

In short, the data type to generate is matrix or vector, and the operations are vectorized. Since the CM language is specialized in matrix/vector types and operations, we choose to generate CM code. 

\subsubsection{Optimized Code Generation for Special Cases}
\label{sec:stt-special-cases}

Above we have illustrated a space-time transformed IR in general. Here we discuss special cases for which we may reduce register usage and/or produce more compact code. 

In the Code 1 (original IR example) above, we can see that a previous value of variable \TT{V}, $V[\vec{z} - \vec{e}]$, is used, and it will never be used again. Therefore, the new value of variable \TT{V}, $V[\vec{z}]$, can reuse the register assigned to the previous value. This is called {\it register coalescing}. Therefore, with some slight change to the code generation pattern, the compiler can generate the following code for space-time transform instead:

\begin{lstlisting}[
    language=C,numbers=left,stepnumber=1,showstringspaces=false,tabsize=1,breaklines=true,escapechar=`]
        $\mathbf{Code\ 3:\ Space-time\ Transformed\ IR\ Example}$
                     $\mathbf{\ after\ Register\ Coalescing}$
    int $V$[extent of $\mathbf{P}\vec{z}$][$\vec{s}^T\vec{e}$];    
    for t = min($\vec{s}^T\vec{z}$); t <= max($\vec{s}^T\vec{z}$); t++
     vectorize for each PE $\vec{x} \in \mathbf{P}\vec{z}$
       $\vec{z} = h(\vec{x}, t)$
       int $tmp$;
       if $\vec{z}$ is in the original iteration space
          $tmp = f(V[\vec{x} + \mathbf{P}\vec{e}][\vec{s}^T\vec{e}]) + g(\vec{z})$
       $V[\vec{x}][1, 2, ..., \vec{s}^T\vec{e}$ - 1] = $V[\vec{x}][0, 1, ..., \vec{s}^T\vec{e}$-2]
       $V[\vec{x}][0] = tmp$
\end{lstlisting}

Note that the second dimension of \TT{V} is now $\vec{s}^T\vec{e}$: 1 register is saved for each PE.
Also, when $\vec{s}^T\vec{e}$ equals 1, Line 10 above (register shifting) can be removed. And if the programmer has not specified \TT{SpaceTimeTransform::CheckTime} when using the \TT{space\_time\_transform()} primitive, the above code can be further simplified:

\begin{lstlisting}[
    language=C,numbers=left,stepnumber=1,showstringspaces=false,tabsize=1,breaklines=true,escapechar=`]
    $\mathbf{Code\ 4:\ Space-time\ Transformed\ IR\ Example\ with}$
    $\mathbf{\vec{s}^T\vec{e}=1, Register\ Coalescing,\ and\ without\  Checking}$
    int $V$[extent of $\mathbf{P}\vec{z}$];    
    for t = min($\vec{s}^T\vec{z}$); t <= max($\vec{s}^T\vec{z}$); t++
     vectorize for each PE $\vec{x} \in \mathbf{P}\vec{z}$
       $\vec{z} = h(\vec{x}, t)$
       $V[\vec{x}] = f(V[\vec{x} + \mathbf{P}\vec{e}]) + g(\vec{z})$
\end{lstlisting}

For example, for the specification shown at the top half of Fig.~\ref{fig:T2S-GPU-overall-arch}, one may find that all the variables can be register coalesced. Further, variable \TT{X} and \TT{Z} have a dependence with a distance $\vec{e}=\bigl(\begin{smallmatrix} -1\\ 1\end{smallmatrix}\bigr)$ and $\bigl(\begin{smallmatrix} 0\\ 1\end{smallmatrix}\bigr)$, respectively. With the scheduling vector $\vec{s}=\bigl(\begin{smallmatrix} 0 & 1\end{smallmatrix}\bigr)$, we have $\vec{s}^T\vec{e}=1$ for both dependences. Also, \TT{SpaceTimeTransform::CheckTime} is not specified. Therefore, the above Code 4 style of code generation can be applied, which results in the transformed IR as shown at the end of Section~\ref{sec:spec-for-1-d-convolve-SBM}.

\section{Generality and Flexibility: with 1-D and 2-D Convolution as Examples}
\label{sec:convolve}

Our approach is based on UREs and space-time transforms. This makes our approach very general: UREs and space-time transforms are the theoretical foundation of most, if not all, systolic arrays. They have been extensively studied during the past several decades, and have been used to design numerous systolic arrays from various domains~\cite{whySystolic,HugheyPhdThesis91,IntroSystolic}. They are also mathematically simple and rigorous.  

This approach is also flexible: from one systolic design, often with slight changes in its UREs and/or transform matrix, a new systolic design can be created. This opens an avenue for exploring the systolic design space for the best designs.

In this section, we demonstrate the generality and flexibility of our approach with various designs of 1-D and 2-D convolution.  We have shown a design, SBM, for 1-D convolution before in Fig.~\ref{fig:1-d-conv-SBM}, and its UREs and space-time transform in the specificaiton in Fig.~\ref{fig:T2S-GPU-overall-arch}. Here we show  several other designs for 1-D convolution in Table~\ref{table:other-1-d-convolution}. Among them, BSM, FSM, BFS, and FFS were proposed by Jo, Kim and Park for an ASIC implementation by manually transforming dataflow graphs~\cite{JoKimPark2018}. Here we express the same designs, but use UREs and space-time transforms instead (See the descriptions in the table). The  UREs and space-time transforms for each design can be used to replace those in the example specification in Fig.~\ref{fig:T2S-GPU-overall-arch}.

To further show the generality and flexibility, in Table~\ref{table:other-1-d-convolution}, we also present a new design, FBS, for 1-D convolution, with a stride of 1 and 2, respectively. 

All these designs for 1-D convolution involve only one projection. To demonstrate multiple projections, we also look at 2-D convolution, which is defined as below:

\begin{equation}
      Z(c, r) = \sum\limits_{q}\sum\limits_{p}x(c+q, r+p) * w(q, p)
\end{equation}

\begin{figure}
    \centering
    \includegraphics[width=\linewidth]{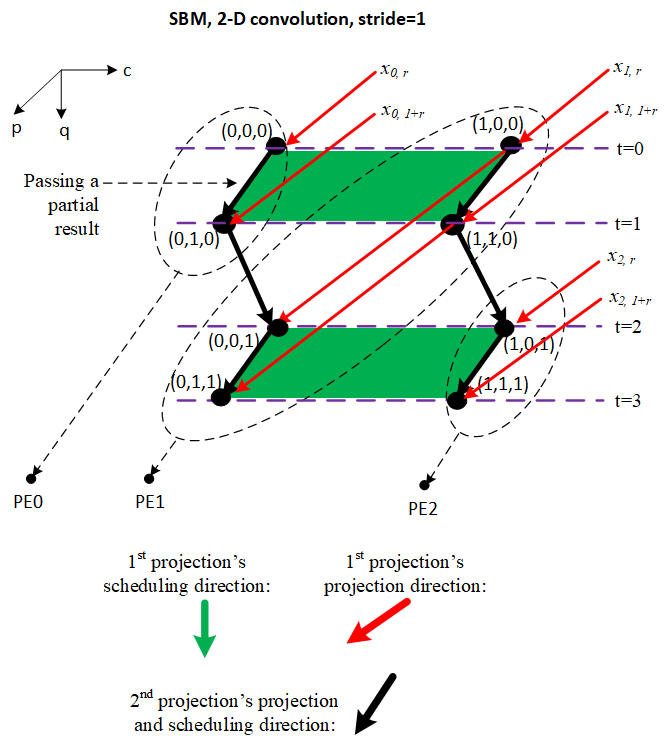}
    \caption{Applying 2 projections to 2-D convolution. In this design, we keep the outer loop \TT{r} untouched, and transform only the inner loops \TT{(c, p, q)}. Every point is annotated with its corresponding indices \TT{(c, p, q)}. For simplicity, the broadcasting of weights is not drawn.}
\label{fig:2-d-conv-SBM}
\end{figure}

Fig.~\ref{fig:2-d-conv-SBM} show the SBM design for 2-D convolution. Although it looks complicated, following the dataflow, we can easily figure out the UREs as follows:
\begin{lstlisting}[
    basicstyle=\ttfamily\footnotesize,language=C,numbers=none,stepnumber=1,showstringspaces=false,tabsize=1,breaklines=true,escapechar=`]
    X(c,p,q,r)=select(c==C-1||q=0,x(c+q,r+p),X(c+1,p,q-1,r))
    W(c,p,q,r)=w(q,p)
    Z(c,p,q,r)=select(p==0 && q==0, 0, 
                select(p==0,Z(c,p+P-1,q-1,r), Z(c,p-1,q,r)))
               + X(c,p,q,r) * W(c,p,q,r)                   
    Out(c, r)=select(p==P-1 && q==Q-1, Z(c,p,q,r))
\end{lstlisting}

Following the two projection directions shown in the figure, we can easily figure out that the allocation matrix is  $\bigl(\begin{smallmatrix} 
  1 & 0 & 1
\end{smallmatrix} \bigr)$, i.e. a computation point at \TT{(c, p, q)} at any given \TT{r} will be mapped to PE \TT{c + q}. 
 
Following the two scheduling directions shown in the figure, we can also easily figure out that the scheduling vector is $\bigl(\begin{smallmatrix} 
  0 & 1 & P
\end{smallmatrix} \bigr)$.

As we can see, UREs and space-time transforms can easily express all these designs. The mathematics involved is intuitive, and should be comfortably mastered by common programmers. Besides, programmers can leverage the wealth of research fruits accumulated in the past several decades. 

\begin{table*}[!htb]
  \centering
  \setlength{\tabcolsep}{20pt}
  \captionsetup{belowskip=8pt,aboveskip=4pt}
  \caption{Several other systolic designs for 1-D convolution. }
  \footnotesize%\small
  \begin{tabularx}{\linewidth}{c c }
    \toprule
    \includegraphics[scale=0.65]{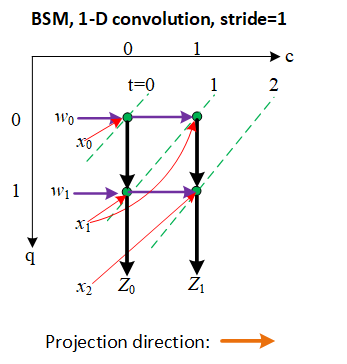}&
    \includegraphics[scale=0.65]{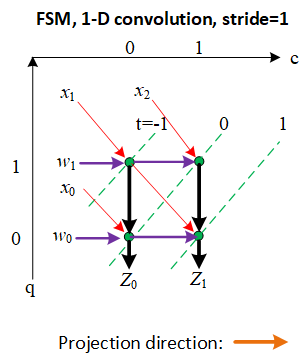}
    \\
    \vtop{\hbox{\strut X(c, q) = x(c + q)}
          \hbox{\strut W(c, q) = select(c == 0, w(q), W(c - 1, q))}
          \hbox{\strut Z(c, q) = select(q == 0, 0, Z(c, q - 1)) + X(c,q)*W(c,q) }
          \hbox{\strut Space-time transform: $\bigl( \begin{smallmatrix} 
                         s \\
                         t
                        \end{smallmatrix} \bigr) = \bigl( \begin{smallmatrix} 
                         0 & 1\\
                         1 & 1 
                        \end{smallmatrix} \bigr) * \bigl( \begin{smallmatrix} 
                         c \\
                         q
                        \end{smallmatrix} \bigr)$}
         }    
 &
    \vtop{\hbox{\strut X(c, q) = select(c == 0 \TT{||} q == Q - 1, x(c + q), X(c - 1, q + 1))}
          \hbox{\strut W(c, q) = select(c == 0, w(q), W(c - 1, q))}
          \hbox{\strut Z(c, q) = select(q == Q - 1, 0, Z(c, q + 1)) + X(c,q)*W(c,q)}
          \hbox{\strut Space-time transform: $\bigl( \begin{smallmatrix} 
                         s \\
                         t
                        \end{smallmatrix} \bigr) = \bigl( \begin{smallmatrix} 
                         0 & 1\\
                         1 & -1 
                        \end{smallmatrix} \bigr) * \bigl( \begin{smallmatrix} 
                         c \\
                         q
                        \end{smallmatrix} \bigr)$}
         } \\
\\
    \includegraphics[scale=0.65]{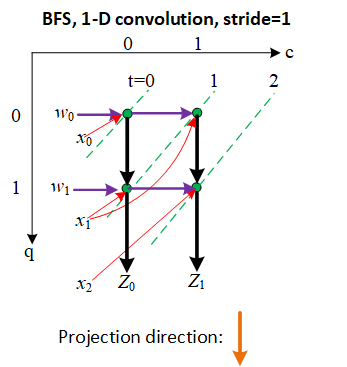}&
    \includegraphics[scale=0.65]{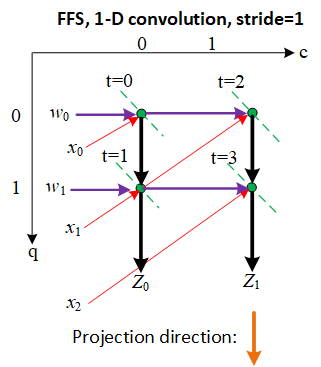}\\
    \vtop{\hbox{\strut X(c, q) = x(c + q)}
          \hbox{\strut W(c, q) = select(c == 0, w(q), W(c - 1, q))}
          \hbox{\strut Z(c, q) = select(q == 0, 0, Z(c, q - 1)) + X(c,q)*W(c,q)}
           \hbox{\strut Space-time transform: $\bigl( \begin{smallmatrix} 
                         s \\
                         t
                        \end{smallmatrix} \bigr) = \bigl( \begin{smallmatrix} 
                         1 & 0\\
                         1 & 1
                        \end{smallmatrix} \bigr) * \bigl( \begin{smallmatrix} 
                         c \\
                         q
                        \end{smallmatrix} \bigr)$}
         } &
    \vtop{\hbox{\strut X(c, q) = select(c == 0 \TT{||} q == Q - 1, x(c + q), X(c - 1, q + 1))}
          \hbox{\strut W(c, q) = select(c == 0, w(q), W(c - 1, q))}
          \hbox{\strut Z(c, q) = select(q == 0, 0, Z(c, q - 1)) + X(c,q)*W(c,q)}
           \hbox{\strut Space-time transform: $\bigl( \begin{smallmatrix} 
                         s \\
                         t
                        \end{smallmatrix} \bigr) = \bigl( \begin{smallmatrix} 
                         1 & 0\\
                         2 & 1 
                        \end{smallmatrix} \bigr) * \bigl( \begin{smallmatrix} 
                         c \\
                         q
                        \end{smallmatrix} \bigr)$}
         }\\
    \includegraphics[scale=0.6]{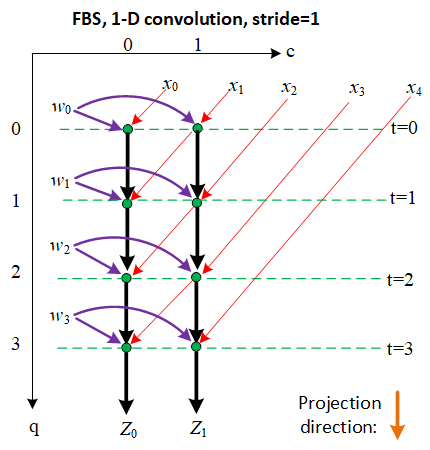}&
    \includegraphics[scale=0.6]{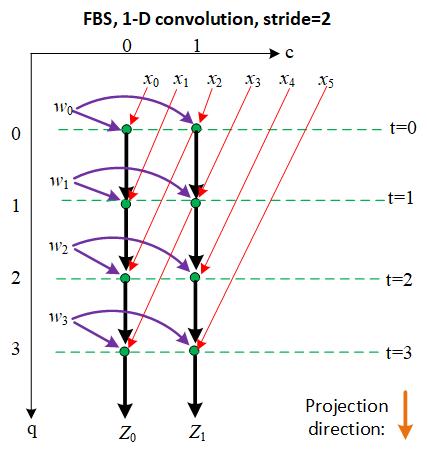}\\
        \vtop{\hbox{\strut X(c, q) = select(q == 0 || c == C - 1, x(c + q), X(c + 1, q - 1))}
          \hbox{\strut W(c, q) = w(q)}
          \hbox{\strut Z(c, q) = select(q == 0, 0, Z(c, q - 1)) + X(c,q)*W(c,q)}
           \hbox{\strut Space-time transform: $\bigl( \begin{smallmatrix} 
                         s \\
                         t
                        \end{smallmatrix} \bigr) = \bigl( \begin{smallmatrix} 
                         1 & 0 \\
                         0 & 1
                        \end{smallmatrix} \bigr) * \bigl( \begin{smallmatrix} 
                         c \\
                         q
                        \end{smallmatrix} \bigr)$}
         } &
    \vtop{\hbox{\strut X(c, q) = select(q < 2 || c == C - 1, x(c + q), X(c + 1, q - 2))}
          \hbox{\strut W(c, q) = w(q)}
          \hbox{\strut Z(c, q) = select(q == 0, 0, Z(c, q - 1)) + X(c,q)*W(c,q)}
           \hbox{\strut Space-time transform: $\bigl( \begin{smallmatrix} 
                         s \\
                         t
                        \end{smallmatrix} \bigr) = \bigl( \begin{smallmatrix} 
                         1 & 0 \\
                         0 & 1
                        \end{smallmatrix} \bigr) * \bigl( \begin{smallmatrix} 
                         c \\
                         q
                        \end{smallmatrix} \bigr)$}
         }         
         \\
\bottomrule
  \end{tabularx}
  \label{table:other-1-d-convolution}
  %\vspace{-2mm}
\end{table*}

\section{Experiments}
\label{sec:experiments}

% Gen9 GPU, there are 24 EUs, and each EU contains 7 EU threads; each EU thread contains SIMD 8/16 lanes for 32/16-bit floats/integers, and 128 vector registers; each vector register has 32 bytes. Programmers should consider these machine-specific constants to decide the best parameters for their specifications, such as the loop splitting factors \ttt{CC} and \ttt{CCC} in the example specification in Fig.~\ref{fig:T2S-GPU-overall-arch}. Our compiler follows programmers' specifications and generate code that is correct and as efficient as possible. 

% ~\footnote{In an Intel Gen9 GPU, there are 24 EUs, and each EU contains 7 EU threads; each EU thread contains SIMD 8/16 lanes for 32/16-bit floats/integers, and 128 vector registers; each vector register has 32 bytes. Programmers should consider these machine-specific constants to decide the best parameters for their specifications, such as the loop splitting factors \ttt{CC} and \ttt{CCC} in the example specification in Fig.~\ref{fig:T2S-GPU-overall-arch}. Our compiler follows programmers' specifications and generate code that is correct and as efficient as possible the SIMD function units are FPUs (floating-point units), which actually support both floating-point and integer computations.}

We have prototyped our approach on top of Susy~\cite{Susy}, generating CM code for Intel GPUs~\footnote{The current prototype has the following engineering limitations, which do {\it not} affect the validity of the experimental results we show in this section, and we are working to remove the limitations soon: (1) Our compiler can successfully generates CM kernel code for the GPU device and x86 binary for the CPU host, but does not have a runtime system yet to connect the host and device code. Thus we have to manually write host code to run a kernel for testing correctness and performance. However, this does not affect the measurement of performance, which is purely the kernels' execution time on the GPU device. (2) The compiler has a bug in generating vectorized code with control flow. We have to manually fix that in the generated code with a tiny code change.  }. In this section, we evaluated the performance of generated code that implements the above 1-D and 2-D convolution designs described in Section~\ref{sec:spec-for-1-d-convolve-SBM} and ~\ref{sec:convolve}. The first SBM design is programmed using about 10 to 20 minutes, and with slight modification of the UREs and space-time transform of the SBM design, all the other designs are quickly expressed in a few minutes. This demonstrates the generality and flexibility of our approach, and the productivity of exploring the design space.

Our testing machine has an Intel HD Graphics 620 GPU, which is manufactured in 14 nm process based on Gen9.5 Kaby Lake architecture, runs at 1.1 GHZ and shares 3.9 GB memory with an Core i5 CPU. This GPU has 3 subslices; every subslice has 8 EUs, and every EU has 7 threads; every EU thread contains SIMD 8/16 lanes for 32/16-bit floats/integers, and 128 vector registers; each vector register has 32 bytes. EU threads in the same subslice can communicate through shared memory.

This GPU has hardware support for 1- and 2-D convolution of output size 16*1 and 16*4. Every subslice of the GPU has a {\it sampler}, which fetches data from external memory, and performs operations including convolution. The CM language has two interface functions,  \TT{cm\_va\_1d\_convolve} and \TT{cm\_va\_2d\_convolve}, to allow programmers to use the hardware samplers to perform convolutions directly.

We issue over 10 millions of threads for sufficiently long and stable execution time, 56 threads as a group, each thread running a convolution (1-D or 2-D, performed by the sampler or our software systolic arrays, with filter size equal to 2 or 5 for 1-D convolution or 2x2 or 5x5 for 2-D convolution, with stride equal to 1 or 2), and every thread generating a 16*4 output matrix. Then we calculate the {\it throughput} by the number of output data divided by the execution time. We report the {\it relative performance} of our systolic arrays by their throughputs divided by the throughputs of the samplers for the same convolutions. We choose to compare with the sampler since it is a specialized hardware, and thus sets up a high-performance baseline. 

\subsection{Summary of the Experimental Results}

Briefly, our experiments yield the following results:

\begin{enumerate}
    \item All our systolic designs work correctly with reasonable performance. Among them, SBM and FBS show the best  performance: they are 47\%-59\% faster than the samplers for 1-D convolution, and are close to or faster than the samplers for 2-D convolution with a small filter size. Profiling shows this is due to their efficient usage of the SIMD lanes in the EUs. Qualitative analysis also confirms that they are the best designs.
    \item The sampler-performed convolutions are limited in flexibility since the filter size can not exceed 8, and the stride is fixed to 1. In comparison, our software systolic arrays are not limited to any specific filter size or stride.
\end{enumerate}

\REM{
We specify the UREs and space-time transforms as shown in Section \ref{sec:convolve}, and the expected CM code is presented in Figure \ref{fig:1d-conv-cm-code}. On account of our compiler immaturity, some situations still cannot be handled efficiently. Take $X(c, q)$ in the FBS design as an example. It requires specializing the last SIMD lane to step-loading a new element from memory, which causes all the elements processed sequentially by the current compiler. To eliminate the inefficiency mentioned above, we manually modify the generated code but keep the code skeleton unchanged, which can still reflect the code's quality.
}

\begin{table*}[!htb]
\caption{1-D Convolution Experimental Results.}
\label{table:1dconv}
\begin{tabular}{|l||c|c|c|c|c|c|c|c|c|c|c|c|}
\hline
 & \multicolumn{2}{c|}{\BF{SBM}} & \multicolumn{2}{c|}{\BF{BSM}} & \multicolumn{2}{c|}{\BF{FSM}} & \multicolumn{2}{c|}{\BF{BFS}} & \multicolumn{2}{c|}{\BF{FFS}} & \multicolumn{2}{c|}{\BF{FBS}} \\ \hline\hline
\BF{Filter size} & 2 & 5 & 2 & 5 & 2 & 5 & 2 & 5 & 2 & 5 & 2 & 5 \\ \hline
\BF{Relative performance} & \BF{157\%} & \BF{147\%} & 67\% & 49\% & 56\% & 39\% & 56\% & 48\% & 33\% & 30\% & \BF{159\%} & \BF{157\%} \\ \hline
\end{tabular}
\end{table*}

\begin{table}[!htb]
\caption{ Analytical Model for 1 1-D Convolution (filter size=5)}
\label{table:1dconv-model}
\begin{tabular}{|l||c|c|c|c|c|c|}
\hline
 & \BF{SBM} & \BF{BSM} & \BF{FSM} & \BF{BFS} & \BF{FFS} & \BF{FBS} \\ \hline\hline
\BF{Outturn} $^{\rm a}$ & \BF{3.2} & 0.8 & 0.8 & 0.8 & 0.46 & \BF{3.2} \\ \hline
\BF{Utilization} $^{\rm b}$ & \BF{80\%} & 80\% & 80\% & 25\% & 14\% & \BF{100\%} \\ \hline
\BF{Reg usage} & 80 & 25 & 35 & 80 & 112 & 80 \\ \hline
\end{tabular}
{\raggedright $^{\rm a}$ Equals 16/(\#time steps) \par}
{\raggedright $^{\rm b}$ Equals 16* (filter size)/(\#SIMD lanes * \#time steps) \par}
\end{table}

\begin{table}[!htb]
\caption{2-D Convolution Experimental Results}
\label{table:2dconv}
\begin{tabular}{|l||c|c|c|c|}
\hline
        & \multicolumn{2}{c|}{\BF{SBM}} & \multicolumn{2}{c|}{\BF{FBS}}          \\ \hline\hline
\BF{Filter size} & 2x2 & 5x5 & 2x2 & 5x5 \\ \hline
\BF{Relative performance}& \BF{92\%} & 20\% & \BF{114\%} &24\%\\ \hline
\end{tabular}
\end{table}

\subsection{Detailed Results}

Table~\ref{table:1dconv} shows the performance of our designs for 1-D convolution. SBM and FBS with a 5x5 filter show outstanding performance among all the other designs. All the other designs show inferior performance.

The difference in performance among these designs is expected. We can derive an analytical model for the designs shown in Table \ref{table:1dconv-model}. To be simple, consider 1 convolution with filter size of 5. Every design outputs 16 results. Thus the average {\it outturn} is 16/(\#time steps). Furthermore, a design performs (\#SIMD lanes * \#time steps) number of operations in a convolution, but  among them only 16 * (filter size) number of operations contribute to useful results, and thus the {\it utilization rate} of the SIMD lanes equals 16 * (filter size) / (\#SIMD lanes * \#time steps). Finally we can count the number of scalar registers the compiler allocates for each design (A vector register is counted as multiple scalar registers). A design with high outturn, high utilization of SIMD lanes, and preferably lower register usage, is a good design, and expected to yield high performance. As we can see from the analytical mode in Table \ref{table:1dconv-model}, SBM and FBS excel at both outturn and utilization, while the other designs are inferior.

\REM{We could observe that both FSM and BFS have a smaller throughput requiring much time to complete. For the FFS and FBS design, not only low throughput but also low utilization they have, only a small number of SIMD lanes do the actual computation. Not surprisingly, they consume more time.
Among the six designs, the BSM, FBS could achieve about 90\% relative execution time. In the absence of manual crafted optimization, the reported performance holds a promise of chasing or even surpassing the expert's implementation. In contrast, the FSM, BFS and FFS could achieve only about 60\% performance.}

Table \ref{table:2dconv} reports the performance of 2-D convolution for filter size of 2x2 and 5x5 with stride of 1. They demonstrate multi-projection and  generality of our approach.  The SBM design has been shown in Fig.~\ref{fig:2-d-conv-SBM}; the FBS design is similar, and simply converts a 2-D convolution into multiple FBS 1-D convolutions. We will work on the other designs of 2-D convolution in future.

As we can see, the two designs are close to or surpass the performance of the samplers for 2-D convolutions when the filter size is small, but the performance drops sharply as the filter size increases. As there is no public document on how the samplers implement 2-D convolution, we have difficulty in understanding the performance difference and will contact Intel experts for help. 

\REM{This property leads to a much worse relative execution time, and we think it persists as an intrinsic limitation of direct convolution since it must iterate the whole filters. Currently, the FBS implementation has more memory pressure than the SBM, which worsens the whole performance, but the computation performance is a little better with the higher utilization. 

Note that 2-D convolution needs to load a matrix into local memory that posed significant memory pressure, and the data loaded by sequential threads have considerable overlaps. Our compiler supports specifying data storage location: shared memory or register. However, optimizations to maximize memory performance are still left as manual work.}

\section{Related Work}
\label{sec:related}

Systolic arrays are usually built on spatial architectures like FPGAs and CGRAs, or built as ASIC circuits~\cite{whySystolic,FPGAImageProcessing,Hrabovsky:2dConvolver:17,Buyukkurt08compilergenerated,settle2013high, TPU}.

The Alpha language~\cite{Verge:1991:alphaLang} is a functional language to express and transform iterative algorithms, and generate systolic arrays. Programmers have limited control on optimizations.
\REM{except the original recurrence equations. 
The MDFL language enables programmers to focus on expressing wavefronts~\cite{Kung:1982:WaveArrayProc}, which greatly simplifies the description of a systolic algorithm. However, the language is for a specially designed wavefront array processor.}

T2S~\cite{Rong:2018:T2S:arxiv} is a methodology that addresses high-performance high productivity programming on spatial architectures like FPGAs.
T2S advocates a programming style that separates concerns by separating a temporal definition from a spatial mapping. 
\REM{A programmer specifies a temporal definition and a spatial mapping. The temporal definition defines the functionality to compute, while the spatial mapping defines how to decompose the functionality and map the decomposed pieces onto a spatial architecture. The specification precisely controls a compiler to actually implement the loop and data transformations specified in the mapping. } T2S-Tensor~\cite{T2SFCCM19} is the first system that realizing the principle of T2S, built upon the Halide language and compiler framework~\cite{Ragan-Kelley:2013:HLC:2491956.2462176}. T2S-Tensor creates asynchronous (i.e. wavefront) arrays, and has demonstrated impressive productivity and performance for  dense matrix multiply, MTTKRP, TTM and TTMc  on an Arria-10 FPGA and a research CGRA. However, T2S-Tensor relies on loop unrolling, which permits creating only a limited set of systolic arrays.  

Susy~\cite{Susy} introduces UREs and space-time transforms into T2S-Tensor, and generates synchronous arrays on FPGAs. It achieves  performance for matrix multiply similar to that of T2S-Tensor, but with much less block RAM usage: thanks to space-time transform, the register requirement of a recurrent variable can be precisely calculated. The limitation of Susy is that it allows only a single projection along a loop dimension.

\REM{
The approach proposed in this paper further extends Susy from FPGAs to GPUs, and allows multiple projections with arbitrary projection directions. GPUs are essentially different from FPGAs: unlike FPGAs that have massive amount of fine-grain logic resources, and have distributed memory over a planar area but without a hardware caches, GPUs have many cores or execution units, have a memory hierarchy with hardware cache(s). 
}

GPUs are usually programmed in SIMT style in CUDA or OpenCL. Our approach takes advantage of the fact that the underlying execution units actually work in SIMD fashion, and therefore, we enable a mixed style of SIMT + SIMD.
\REM{Programmers can express multiple threads in terms of groups and threads,  and within a single thread, express a linear systolic array that working in SIMD style. }

Our approach leverages Susy, breaks its limitations and extends it to GPUs, with substantial innovation as summarized in Section~\ref{sec:intro}. We are not aware of any other language and compiler  for creating  systolic arrays on GPUs.

\REM{GPUs are usually programmed in SIMT style in CUDA or OpenCL, and programmers have to write explicit data shuffling, loop transformations, etc. to create a systolic array~\cite{WangjieSequenceAlignment}.  
We point out that UREs and space-time transforms are our unique contributions to the original Halide system. Halide could be used to generated on GPUs some SIMD codes that are equivalent to certain special systolic arrays, but essentially Halide does not have the capability to express arbitrary systolic arrays. In fact, Halide does not allow recurrent functions.}

\section{Conclusion and Future Work}
\label{sec:conclusion}

We have presented a programming language and compiler, for the first time to the best of our knowledge, for quickly building high-performance systolic arrays on GPUs. Based on UREs and space-time transform, our approach has a solid mathematical foundation and should be able to cover a wide range of applications. Primitive experimental results on Intel GPUs have confirmed the generality and flexibility of this approach, and demonstrated its promise in performance and productivity.

In future, we will build a design space explorer that can automatically enumerate valid systolic designs for a workload, and search for the best candidates, and we will extend the system to CPUs as well, since CPUs also have SIMD execution units.

%% Acknowledgments
\begin{acks}                            %% acks environment is optional
                                        %% contents suppressed with 'anonymous'
  %% Commands \grantsponsor{<sponsorID>}{<name>}{<url>} and
  %% \grantnum[<url>]{<sponsorID>}{<number>} should be used to
  %% acknowledge financial support and will be used by metadata
  %% extraction tools.
Size Zheng, Xiuping Cui, and Yunshan Jia have explored systolic computing on GPUs from the perspective of vectorization. Anand Venkat has studied systolic implementations of matrix multiply on GPUs, with generous help from Sabareesh Ganapathy. We appreciate their efforts even though we did not follow their technical paths. Fangwen Fu and Evghenii Gaburov discussed with us some interesting challenges of programming GPUs. Kai Yu Chen, Guei-Yuan Lueh, and Gang Y Chen kindly helped us understand the CM language. Christopher J Hughes and Geoff Lowney have provided solid support for our research.

\end{acks}

\clearpage
\bibliography{local}

\end{document}